\documentclass[Afour,sageh,times]{sagej}

\usepackage{moreverb,url}
\usepackage{natbib}

\usepackage[colorlinks,bookmarksopen,bookmarksnumbered,citecolor=red,urlcolor=red]{hyperref}

\newcommand\BibTeX{{\rmfamily B\kern-.05em \textsc{i\kern-.025em b}\kern-.08em
T\kern-.1667em\lower.7ex\hbox{E}\kern-.125emX}}

\def\volumeyear{2023}

\begin{document}

\runninghead{What is the disinformation problem?}

\title{What is the disinformation problem? Reviewing the dominant paradigm and motivating an alternative sociopolitical view}
\author{Nicholas Rabb}

\affiliation{\affilnum{1}Tufts University, US}

\corrauth{Nicholas Rabb
Tufts University,
Medford, MA 02145,
US.}

\email{nicholas.rabb@tufts.edu}

\begin{abstract}

Disinformation research has proliferated in reaction to widespread false, problematic beliefs purported to explain major social phenomena. Yet while the effects of disinformation are well-known, there is less consensus about its causes; the research spans several disciplines, each focusing on different pieces. This article contributes to this growing field by reviewing prevalent U.S. disinformation discourse (academic writing, media, and corporate and government narrative) and outlining the dominant understanding, or paradigm, of the disinformation problem by analyzing cross-disciplinary discourse about the content, individual, group, and institutional layers of the problem. The result is an individualistic explanation largely blaming social media, malicious individuals or nations, and irrational people. Yet this understanding has shortcomings: notably, that its limited, individualistic views of truth and rationality obscures the influence of oppressive ideologies and media or domestic actors in creating flawed worldviews and spreading disinformation. The article then concludes by putting forth an alternative, sociopolitical paradigm that allows subjective models of the world to govern rationality and information processing -- largely informed by social and group identity -- which are being formed and catered to by institutional actors (corporations, media, political parties, and the government) to maintain or gain legitimacy for their actions.

\end{abstract}


\maketitle

\section{Introduction}

The threat of disinformation and misinformation to democracy is a growing concern in nations across the world. Both in the mainstream discourse and in academic research, many are investigating the ways that flows of information are leading to social tension because of political disagreement \citep{iyengar2009red,sunstein2001Republic}, hateful discourse \citep{sobieraj2020credible,boatright2019crisis}, and of beliefs that lead to harmful behaviors \citep{pewAsianAmericanViolence,pewMediaCovidExaggeration} -- which is traced back to information that is manipulative and not true.

While the \emph{effects} of the disinformation crisis are being felt across the globe and agreed upon, the actual \emph{problem} -- what is causing and continuing the crisis -- has less consensus. Even just in the U.S., there are many accounts of what causes disinformation that, when analyzed together, can become contradictory and confusing. Yet a reading of discourse surrounding U.S. disinformation makes clear that there is a prevailing narrative describing the problem -- one described in highly-cited academic research, mainstream news media, and other forms of media such as film -- and other, less popular descriptions. But this narrative is scattered between individual disciplines (computer science, network science, media studies, psychology and cognitive science, political theory) and only takes form when looking at the whole.

An account of the ``disinformation problem'' can be characterized by arguments organized into a multi-layer framework that examines the \emph{content} of disinformation, its effects on \emph{individuals}, \emph{groups}, and its articulation through \emph{institutions}. The culmination of arguments across these layers reveals a narrative describing the problem, a paradigm \citep{kuhn1970structure} that in turn informs what problems are deemed relevant and what research is performed.

A synthesis of a breadth of literature reveals the \emph{dominant paradigm} to have a content layer that is primarily concerned with clearly distinguishing truth from falsehood in a quest for objectivity; an individual layer where disinformation believers are painted as irrational, ``stupid,'' or conspiratorial thinkers; a group layer concerned with echo-chambers and polarization as a key consequence and driver of disinformation; and an institutional layer where individual rogue actors take the spotlight, and if any institutions are named as perpetrators, they are typically social media companies or foreign nations. 

Yet the dominant paradigm has theoretical and empirical flaws stemming from its focus on objective truth and attribution of the problem to social media and its manipulation by malicious or foreign actors. Individuals believe certain information because of its sensibility or resonance within their own subjective minds and worldviews; fundamentally formed on top of sociopolitical ideologies and containing differing levels of trust in institutional or individual bearers of truth. Some information is believed because it satisfies social or psychological needs. The overemphasis on social media as the originator of the problem is based on early speculative research that was disseminated widely and uncritically by media outlets, and disregards that these sociopolitical belief processes have always been at play, have been manipulated by political actors throughout history, and are most often hijacked by domestic powerbrokers including the government, political parties, and corporations.

The critique of the dominant paradigm springboards to an \emph{alternative sociopolitical paradigm} that could be used to better understand the causes of disinformation. This view makes clear the need to identify sociopolitical context in disinformation media: ideological underpinnings, in- or out-group dynamics, position signaling, motivated reasoning arguments, and trust cues. The same can be done for individuals and groups who interact with this content, as their belief of certain messages depends on how the sociopolitical content matches their worldviews. Disinformation messaging perpetuated by institutions should be viewed in the additional context of their political economic incentives; whether they be governments, political parties, corporations, or other organizations.

The resulting view concludes that the disinformed are not simply irrational, but operating rationally within flawed or manipulated worldviews (which themselves may be the product of disinformation) constructed by institutions playing to social and psychological triggers. False and problematic information that appeals to these worldviews, or that plays on certain trust cues, can be believed by virtue of its resonance with existing beliefs. Continued belief in disinformation may serve social or political purposes such as maintaining group membership, or be the result of dissonance reduction processes, as questioning one's worldview and identity can be psychologically painful. Flawed beliefs may not be able to be addressed without deeply understanding the subjectivity and worldviews of the disinformed, and crafting informational campaigns that resonate but also correct gradually to more productive or true beliefs.

A sociopolitical view of disinformation demands new research techniques and directions, analyzing past and present campaigns to understand the subjectivities and worldviews at play, their interactions with media information, and the major institutional actors. This work lays the foundation for a turn in disinformation research that could lead to deeper understandings of why there is widespread belief in false and problematic information, and make strides towards addressing it and moving to a more reasonable and democratic society.

\section{Scope and Scale}

\subsection{What is the Disinformation Problem?}

This paper is concerned with \emph{disinformation}, as opposed to \emph{misinformation} or \emph{fake news}. Though the definitions are not agreed upon, there have been theoretical distinctions made between the three: disinformation as intentional manipulation for political or economic gain, misinformation as false information unintentionally spread, and fake news as information which literally attempts to mimic news formats to trick audiences \citep{benkler2018network,marwick2018people}.

To understand why political discourse is vitriolic \citep{sobieraj2020credible,boatright2019crisis} and confusing, why disbelief in life-saving public health measures like Covid-19 vaccination \citep{pewTwentyFindings,pewPeristentDifferences} or mask-wearing is widespread, or why there exists such difficulty in reaching cohesion as a society \citep{pewAmericansTrust}, a complicated argument must be put forward. This type of analysis has been called for by scholars of disinformation \citep{kuo2021critical,anderson2021propaganda,marwick2021critical}, who outline the need for disinformation study that considers its history, its disparate impacts on different communities, the harmful nature of the content itself, and the role of institutions in propagating disinformation.

Importantly, this analysis will be U.S.-centric. As a result, the final analysis being limited, but also allows for a deeper dive into a specific context. Systems of disinformation in the U.S., and the resulting political conditions, are complicated in and of themselves. Moreover, the U.S. has a long history of disinforming campaigns that are worth understanding and applying even to nation states other than the U.S. \citep{carey1997taking,chomsky1994manufacturing}. Similar analyses to this could be performed for other nations to understand their context, and the constellation of forces creating problematic disinformation, yet this paper leaves that to future work.

\subsection{Layers of Analysis}

This analysis will utilize a framework that allows for examining a problem from multiple layers of social organization. Similar work has been done by scholars describing phenomenon as in Hill-Collins' Matrix of Domination \citep{collins2002black}, or in Marwick's sociotechnical analysis of disinformation \citep{marwick2018people}.

\begin{figure}
    \centering
    \includegraphics[width=\linewidth]{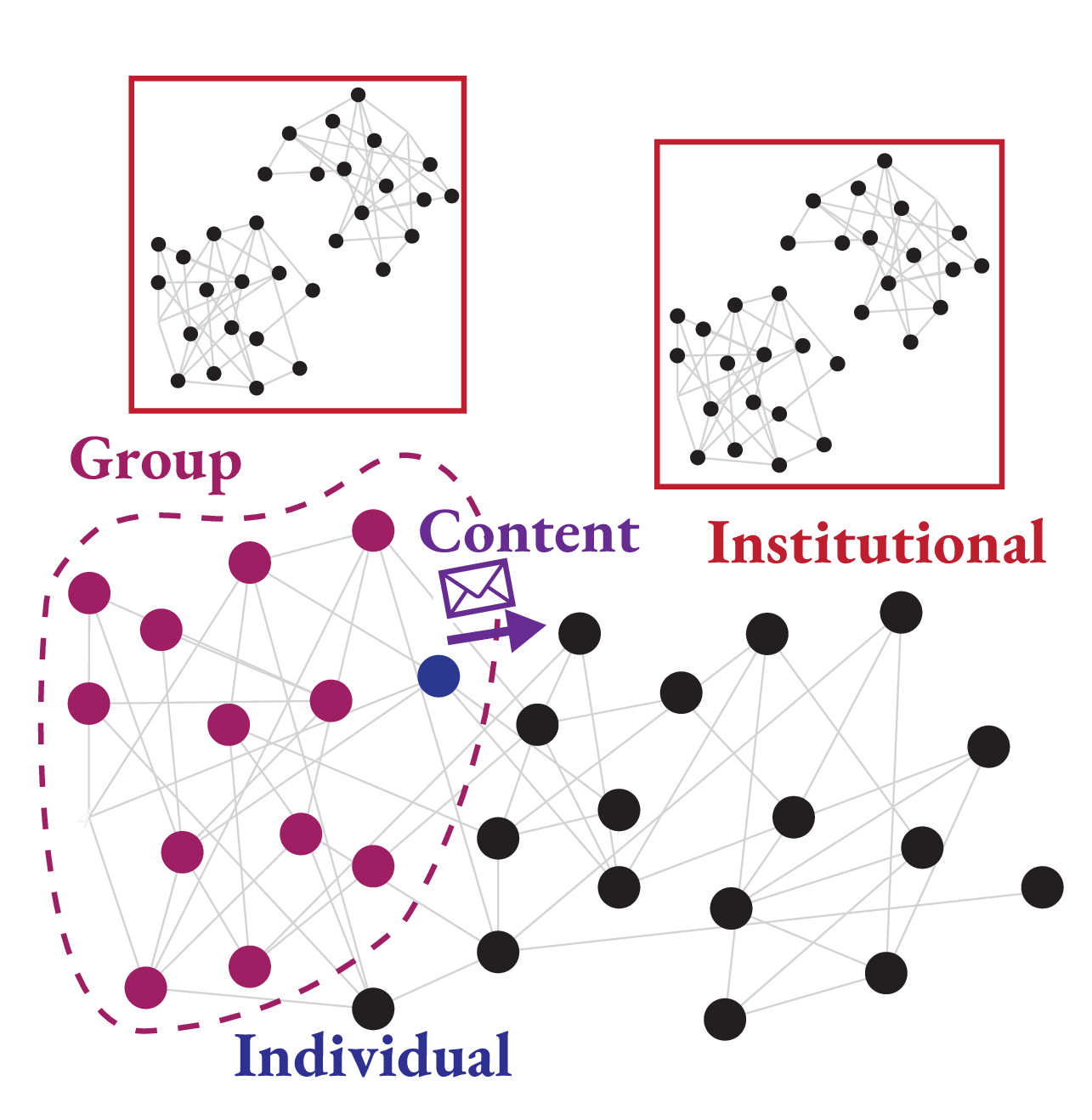}
    \caption{An illustration of the four layers of a paradigm analyzed in this article: content, individual, group, institutional.}
\end{figure}

The framework will view the problem through the \emph{content}, \emph{individual}, \emph{group}, and the \emph{institutional} layers. These four facets of the disinformation problem manifest different mechanisms, and work distinctly from, but in connection with, each other. Disinformation is a highly interdisciplinary topic of study because it operates on so many different layers. Each of them must be understood if we are to have an accurate picture of the problem. The culmination of each of these layers constitutes a \emph{paradigm} -- an often unspoken understanding of the problem that subsequently shapes research and the questions that motivate it \citep{kuhn1970structure}.

\section{Dominant Paradigm of Disinformation}

The dominant paradigm of disinformation spans the layers described above: content, individual, group, and institutional. Each layer of this dominant view can be described by examining popular literature and discourse that focuses on that area of the problem. Each layer shares similar assumptions about disinformation that mirror some features of dominant U.S. society; namely individualism, belief in objectivity, a mythic past free of disinformation, and seeing its cause in foreign powers rather than domestic actors.

Yet the dominant paradigm is not without faults. After laying out the dominant view at each layer, it will be critiqued. Many claims put forth by the dominant paradigm do not hold up given some critical scrutiny. This critique will serve to deconstruct the dominant paradigm so a new paradigm can be motivated in its place.

\subsection{The Content Layer}

\subsubsection{Provably true and false}

Understandably, many conceptions of the disinformation problem focus on truth and falsehood. One embodiment of this comes in the form of advocacy for fact-checking. For example, after multiple waves of Covid-19 disinformation, researchers argued that fact-checking health information would serve a role in successful public health measures \citep{swire2020public}. The World Health Organization (WHO), in response to Covid disinformation, also turned to fact-checks, warning against a simultaneous  ``infodemic,'' and mobilized resources on their website to provide fact-checks on public health myths and to debunk rumors \citep{zarocostas2020fight}.

This desire to rend truth from falsehood at scale is also embodied in algorithmic interventions into the disinformation problem. Technology companies argue that human fact-checking is difficult to perform at the scale of, for example, social media platforms. They argue that computational algorithms can meet the scale of the problem \citep{facebook2019nextPhase,facebook2021how}. This has led computer science researchers to develop machine learning classifiers that attempt to learn patterns that distinguish true messages from false ones \citep{varma2021systematic}.

\subsubsection{Striving for objectivity}

The dominant paradigm's focus on truth and falsehood has a corollary in the form of arguments towards the pursuit of objectivity. This discourse has roots in U.S. journalistic norms for media organizations, but it bleeds into the domain of other institutions who spread information.

The idea of objectivity is deeply embedded in U.S. media culture. As journalism professionalized, there was a movement to turn from the typical partisan reporting of the early 19th century to a more ``detached, observational writing'' \citep{meyerAbusesofObjectivity}. After President Trump's attacks on mainstream journalism during his presidency, mainstream newspapers reacted, doubling down on their role as objective truth-tellers. For example, at that time, the Washington Post changed its slogan to ``Democracy Dies in Darkness.'' The reverence of U.S. media objectivity also refers to the freedom of the press that is enshrined in the First Amendment of the U.S. Constitution, or to the media's historical role in incidents like the Watergate scandal, or in the Vietnam War \citep{chomsky1994manufacturing}.

\subsubsection{Truth and objectivity are more complicated}

In one sense, disinformation focusing on simple notions of truth and falsehood obscures some of the key mechanisms behind the problem. Some messages are untrue or simply problematic because of the context that gives them meaning, not the words themselves. Stanley, in his work decoding the mechanisms behind propaganda, demonstrates this with the sentence, ``Muslims are among us'' \citep{stanley2015propaganda}. Syntactically and semantically, it is true, but the intent behind the statement is to induce fear of Muslims. Arriving at that conclusion relies on the social context that the words invoke (negative views of Muslims), and subtleties of rhetoric (the words ``among us'' invoking fear).

Other political speech, such as that from \emph{Tucker Carlson Tonight} on \emph{Fox News}, is false in a more complicated way, because of nuanced logical fallacies. After Roe v. Wade was overturned by the Supreme Court, Carlson argued the following: companies are burdened by workers with children, who make their healthcare plans more expensive, therefore companies are incentivizing abortions to cut costs \citep{carlsonRoevWade}. While the premises may be true (workers with families have higher healthcare costs; some companies promote abortion as a right), the conclusion is not valid. There may be many reasons a company promotes the legality of abortion.

In another, more general sense, truth is not such an easy pursuit. In most cases, it is arduously arrived at by gathering as much evidence as possible and puzzling through it, as is the case in high-quality scientific research or journalism. But key to the epistemic foundation of science, for example, is that what is ``true'' changes over time as continued research finds flaws and improves theories. What is ``true'' is often what is \emph{most true} at a given point of time, not a forever settled matter.

When epistemic authorities like scientists arrive at ``true'' conclusions, they are often the result of complicated, painstaking analysis and experimentation. No person has the time nor ability to verify all claims to truth that they adopt and then behave according to, which necessarily brings in an element of trust. We rely on the methods of epistemic authorities, like scientists and journalists, to bring us the most true conclusions. But often, these conclusions are shown to be wrong either because other scientists put forward better research, or more evidence comes to light for journalists and they must retract previous conclusions.

This creates troubles for the notion of objectivity, as epistemic authorities cannot claim to always be right, but should rather claim to be committed to the most rigorous pursuit of truth. Yet blind spots, biases, and mistakes will always prevail, as will be discussed further in the institutional layer. Claims to objectivity obscure the ways that authorities can be wrong, and then cast those who have valid critiques of epistemic authorities as problematic disbelievers. Slogans like ``believe science'' miss the point of the institution itself, which is to embody skepticism rather than blindly trust authoritative claims. The more that epistemic institutions claim to always be right, the more that their mistakes reduce trust in those institutions, and lead to disbelief in what may truly be the most valid conclusions.

This problem subsequently affects epistemic tools aimed at tackling disinformation like machine learning classifiers and human fact-checkers. Data sets and fact-checkers may be subject to mistakes and biases. Claiming the objectivity of these tools, often critiqued as a ``view from nowhere'' \citep{katz2020artificial,noble2018algorithms}, rather than acknowledging their subjectivity (in the sense of a ``data setting'' \citep{dignazio2020data}), is both false and can lead to distrust of systems when they inevitably fail.

None of this is to say that some things are not more true than others, but rather that reducing the problem to simple or obvious truths ignores the complexity of truth and our reliance on trust for forming our worldview. It also misses the fact that sociopolitical context and worldview shape how words and phrases and interpreted, which will be further described in the sections to come.

\subsection{The Individual Layer}

\subsubsection{Irrational thinking}


There is a tendency in dominant disinformation research to diagnose the rationality of those who believe in false narratives. One popular strand of cognitive research uses measures correlated with rational thinking to argue that disinformation believers are ``lazy'' \citep{pennycook2019lazy} or ``irrational'' thinkers \citep{pennycook2020falls,bronstein2019belief,pennycook2021psychology}. It argues that more thinking and deliberation is needed on the part of those who fall prey to false narratives \citep{bago2020fake}. Their solutions advocate for behavioral interventions that ``nudge'' individual media consumers towards making rational decisions when interacting with news \citep{pennycook2020fighting,guess2020digital}. These studies argue that individuals who believe in disinformation score lower on cognitive measures like ``open-minded thinking,'' and ``cognitive sophistication,'' enabling researchers to scientifically justify calling disinformation believers ``lazy.''

Several psychological studies have similarly tested individuals' level of ``conspiracy thinking'' \citep{uscinski2020people,enders2021misinformation,enders2021relationship}, measured by a validated questionnaire developed by \citep{uscinski2013people} that includes attitudes like ``even though we live in a democracy, a few people will always run things anyway,'' ``the people who really `run' the country are not known to voters,'' and ``big events like wars, the current recession, and the outcomes of elections are controlled by small groups of people who are working in secret against the rest of us.''

The discourse surrounding cognitive deficiency or conspiratorial thinking is mirrored in mainstream media accounting of disinformation believers, albeit in a less nuanced and more judgemental way. Organizations like CNN, MSNBC and the Washington Post have authored articles with disparaging language; calling QAnon believers ``dumb,'' \citep{cnnQAnonDumb,msnbcTheBeat10821,washPoHowDumb,cnnNBAisBack} ``sad,'' \citep{cnnSidneyPowell} and taking sarcastic tones while describing disinformation believers \citep{msnbcHomelandSecurity,cnnQAnonPhysically}. This, in contrast to the academic research, features more directly demeaning language, but still utilizes the same analysis and conclusion: disinformation believers are being irrational.

\subsubsection{Necessary manipulation towards truth}

One conclusion of this thinking is that there is true information that individuals should be nudged towards through platform manipulations or other means. Social media companies have been investigating the potential of their platforms to manipulate users \citep{matz2017psychological,zuboff2019age} and merge it with behavioral economic trends advocating for ``nudging'' to improve social outcomes \citep{pennycook2020fighting,pennycook2021psychology,swire2020public,van2020using}. Companies like Facebook have become notorious for experimenting with informational cues or news feed algorithm manipulations designed to push users away from disinformation \citep{facebook2019nextPhase,matz2017psychological}. Other computer science research even extends into theoretical algorithmic developments, imagining how to make individuals most susceptible to manipulations \citep{abebe2021opinion}, and then target those susceptible people with desirable information.

\subsubsection{Rationality works within subjective epistemology}

As with truth, rationality is also not as simple as it may seem at first glance. The process of forming a coherent worldview is not solely a matter of one's ability to think hard, or how smart they are. Political belief formation involves psychological processes that are deeply emotional in nature, and individuals believe certain things because they may satisfy diverse psychological or social needs \citep{jost2003political,jost2009political}. Different worldviews, formed by the information one has consumed and how it has interacted with their prior worldview and psychology, determine what is rational or irrational. Individual epistemic frameworks, much as those that determine truth as described above, are subjective, beholden to prior experience, social position, and more \citep{marwick2018people,graan2020fake}.

Individuals can be said to be rational inside of \emph{their models of the world} (molded by epistemic, existential, and relational forces, prior experience, etc.). Calling them irrational, under this framework, becomes a character judgment of their mode of reasoning, and feeds into an uncritical conclusion that disinformation is only believed by a class of less smart or crazy people.

Perhaps surprisingly, Marwick showed that those engaging with conspiracies like QAnon are actually embodying admirable civic virtues -- questioning what they hear and engaging in political research -- albeit arriving at problematic conclusions \citep{marwick2020qanon,kuo2021critical}. These individuals are, in fact, not lazy in the slightest. Many of them are engaged citizens operating within an epistemic framework that leads them to incorrect, harmful conclusions.

Empirical evidence is beginning to show that rationality-based interventions through tools designed to manipulate and nudge are not showing strong evidence of efficacy. Benkler, Faris and Roberts' investigation into disinformation around the 2016 election demonstrated much stronger effects from mainstream media television than from manipulations like those of Cambridge Analytica \citep{benkler2018network}. A large-scale post-hoc evaluation of initial theorizing around how to address Covid-19 disinformation similarly revealed that attempts at nudging proliferated after 2019, but had mixed effectiveness \citep{ruggeri2022evaluating}. It seems likely that different interventions, perhaps taking into account subjective worldviews and subsequent epistemic frameworks, are needed.

\subsection{The Group Layer}

\subsubsection{High-choice polarization}

From the early days of the internet, scholars have articulated worries that the shear volume of information available would negatively affect democracy. MIT Media Lab founder Nicholas Negroponte argued in his 1995 \emph{Being Digital} that the internet enabled a threat that he called the ``Daily Me'' \citep{negroponte1997being}, a virtual newspaper that is tailored to an individual's tastes, essentially predicting (or influencing) the advent of recommendation systems and customized newsfeeds creating echo-chambers. Sunstein's popular analysis of emerging internet technologies such as blogs and news search in \emph{Republic.com} \citep{sunstein2001Republic} aggregated psychological and cognitive research to argue that internet-based communities, formed around interest rather than geography, lead to \emph{echo-chambers} and polarization.

This fear of the confirmation bias-to-polarization pipeline became pervasive in popular discourse and the research community, influencing the media studies agenda to investigate these theorized effects. Media studies turned to studying the phenomena of \emph{audience fragmentation} (the partition of media consumers into many small audiences) \citep{webster2012dynamics} and \emph{selective exposure} (the tendency of media consumers to consume messages in a systematically biased manner diverging from the composition of available messages \citep{cardenal2019digital,messing2014selective,karlsen2020high,knobloch2014choice}).

As internet technology developed, the advent of social media drove these fears into overdrive. Social media has been argued to increase fragmentation and the ability to selectively expose oneself to bias-confirming media \citep{arendt2019selective}, and is now popularly believed to facilitate echo-chamber formation and biased information consumption \citep{webster2012dynamics,iyengar2009red,cardenal2019digital,messing2014selective,karlsen2020high,knobloch2014choice,bakshy2015exposure}.

This all has led many media researchers to argue that the disinformation problem, where \emph{polarization} of political opinion is frequently reported \citep{iyengar2009red,bakshy2015exposure}, is being caused by fragmentation, selective exposure, and its resultant echo-chambers. Moreover, it has led to a pervasive emphasis on polarization itself as the problem, sparking others to model polarization pressures or empirically study them to try to figure out how to reduce them \citep{baldassarri2007dynamics,dandekar2013biased,goldberg2018beyond}.

\subsubsection{Echo-chambers and polarization revisited}

The idea that echo-chambers have caused a polarized, disinformation-ridden society is not empirically supported, but may have been uncritically spread in popular discourse by media and influential political figures. Guess shows that popular journals like \emph{The Independent} have claimed that ``Social media echo chambers gifted Donald Trump the presidency''; that \emph{Wired} proclaimed that ``Your Filter Bubble is Destroying Democracy''; and that even President Obama has criticized ``balkanized'' media for effects on democracy \citep{guess2018avoiding}. 

Counter to this view, there is a growing body of evidence that media consumers are exposed to a diversity of information despite social media and recommendation algorithms \citep{benkler2018network,guess2018avoiding,webster2012dynamics}. Generally, people are aware of arguments from the other side of the political spectrum, and are not totally encapsulated from countervailing evidence, but must somehow be choosing to disregard it \citep{guess2018avoiding}.

Historically, different ``realities'' have been manufactured in the press for centuries (e.g. white press portraying Black people as criminals), and are not a novel issue due to technology \citep{kuo2021critical}. Perhaps those now feeling the presence of echo-chambers have never before had their dominant reality challenged. The focus on echo-chambers and polarization as \emph{themselves} being the problem misses the importance of what the actual content inside the communities is, and countering those ideas \citep{garrett2017echo}.

Polarization is also widely taken uncritically to be something that is \emph{de facto} bad. Yet some social scientists and historians studying polarization argue that it arises at times when society is experiencing extreme inequality, and has historically changed problematic societies into better ones \citep{lakey2016viking,putnam2000bowling}. This frames polarization as something that occurs in periods of social inflection, where the outcome could be good or bad. It emphasizes the \emph{relativity} of polarization, where depending on the initial conditions, a polarized population may be more desirable, especially if it leads to a better outcome.

\subsection{The Institutional Layer}

\subsubsection{Rogue actors}

Clickbait factories and bots are a centerpiece of the dominant view of disinformation. Academic research centers their influence in spreading disinformation \citep{ognyanova2021network,swire2020public,benkler2018network}, and this is mirrored in media. In the wake of the 2016 election, Twitter released a data set of foreign accounts, including bots, allegedly used to sway politics through social media \citep{twitter2018EnablingFurther}. This type of data became cited by the Mueller Investigation and discussed widely \citep{benkler2018network,polyakova2019Mueller}. Similarly, a 2016 story about Macedonian teens generating fake news to make substantial advertising revenue caused a panic about clickbait and its threats to a healthy media environment \citep{graan2018mills}. That story was picked up by many news outlets and even discussed by former President Obama in the media. In the following years, popular films like \emph{The Social Dilemma} dramatically depicted the influence of bots on politics to a wider audience \citep{socialDilemma2020}.

Mainstream media institutions also frequently publish stories pointing to low-level celebrities who play a role in spreading false information. When discussing the spread of Covid disinformation online, \emph{The New York Times} published several stories pointing to the influence of individuals like Joseph Mercola \citep{nyt2021MostInfluential}, Joe Rogan \citep{nyt2021AnimalMedicine}, and various scientists, doctors, or MMA fighters \citep{nyt2020Plandemic}. 

\subsubsection{Revisionist powers \& rogue regimes}

Discussions of the influence of bots and rogue technological actors often overlap with fears of the influence of foreign actors: usually Russia, China, Iran and North Korea, sometimes with a nod to Venezuela, Nicaragua, or Cuba. Coincidentally, these nations align with official U.S. government enemies: the ``revisionist powers'' (Russia and China) and ``rogue regimes'' (Iran, North Korea) as declared in the Department of Defense's 2018 National Defense Strategy \citep{dod2018NationalDefense}, and the ``troika of tyranny'' (Venezuela, Nicaragua, and Cuba) as proclaimed by former National Security Advisor John Bolton \citep{nsc2018BoltonLatinAmerican}. Twitter's aforementioned dataset of bots and tweets designed to influence the U.S. 2016 election, for example, featured only Russian, Iranian, and Venezuelan accounts \citep{twitter2018EnablingFurther}.

Much of the focus, however, falls specifically on Russia and their disinformation tactics. In the wake of the 2016 election, there was a tremendous movement to blame the outcome on Russian interference \citep{wp2018facebookAdDump,benkler2018network}, sparking the Mueller probe. This developed into an enormous story, spanning years of coverage in mainstream media and inevitably filtered into popular media as well. \emph{The Social Dilemma} also featured an explanation of Russian-organized protests that allegedly influenced the outcome of the election, and asserts their desire to sow discord in the U.S. generally \citep{socialDilemma2020}. 

\subsubsection{Social media thwarting objective media}

Most of the research cited in previous sections is framed around analyzing the effect of social media on political outcomes. Whether through analyzing information diffusion online \citep{goel2016structural,del2016spreading,bakshy2015exposure,brady2020mad} and how to counter its problems \citep{pennycook2020fighting}, or changes to the media ecosystem introduced by high news availability \citep{iyengar2009red,cardenal2019digital,messing2014selective,webster2012dynamics}, there is simply an enormous amount of social media research pertaining to disinformation. This makes sense, as data may be more easily available than conducting studies on the rest of the media ecosystem, but it also reflects a widespread anxiety about the effects of social media on democracy and politics \citep{kuo2021critical,tufekci2017twitter}.

This focus has understandably spilled into mainstream discourse, as social media companies have come under fire for their role in the spread of disinformation. There are dozens, if not hundreds, of articles written about the spread of conspiracies and false information on platforms like Facebook, Twitter, Instagram, and YouTube; including some high-profile investigations like \emph{The Wall Street Journal's} ``Facebook Files'' \citep{wsj2021FacebookAngrier,wsj2021FacebookVaccinated}, \emph{The Social Dilemma's} castigation of social media by former executives and senior developers \citep{socialDilemma2020}, and \emph{The Great Hack}'s portrayal of Cambridge Analytica's partnership with Facebook to meddle with U.S. elections on behalf of the Trump campaign \citep{greatHack2019}.

All of the attention given to social media as a culprit behind the disinformation problem suggests that the political situation was much better before its ascendance, and that a return to primarily getting news from mainstream journalism institutions is the solution \citep{kuo2021critical,marwick2018people}. As described in the section on content, there is a strong narrative in the U.S. that mainstream journalism institutions are courageous keepers of truth, in the business of speaking truth to power, and key to ensuring the functioning of democracy.

\subsubsection{Diverting from larger culprits}

While the effect of rogue individuals, foreign states, and social media platforms is certainly something to be considered, there have been careful, convincing analyses that conclude that despite the \emph{presence} of these influences, their \emph{effect} is significantly less than others \citep{goel2016structural,benkler2019cautionary}. This focus obscures the influence of other, more powerful institutions that create disinformation -- namely the U.S. government and private corporations -- while simultaneously erasing the long history of those institutions creating disinformation.

Rogue individuals are inevitably actors within, and products of, a larger system. Media organizations often precede what influential individuals say, shaping them through agenda setting (coverage determining what people think is significant), priming (coverage of certain issues making audiences receptive to particular themes), and framing (organizations constructing stories to further a point of view) \citep{benkler2018network,marwick2018people}. Moreover, convincing empirical evidence has shown that large-scale virality from individual sources is rare, and tends to occur when mainstream media covers a story \citep{goel2016structural}.

Regarding bots and clickbait factories, Benkler, Faris and Roberts performed an extensive analysis of the \emph{presence} of fake news clickbait campaigns during the 2016 U.S. presidential election, and their \emph{actual effects}, ultimately finding that effects are insignificant as compared to those of mainstream media organizations \citep{benkler2018network}. They also note that the process of detecting bots is a highly imprecise science: attempting to classify an account as a bot based off of language usage or interaction patterns is likely to fall prey to biases, cultural differences, and other assumptions. It is thus unclear whether research examining bot networks are legitimate evidence of Russian, Iranian or Venezuelan influence, or a manifestation of biases against typical U.S. ``enemies.'' The same analysis holds for analyzing the effects of Russian information operations. Benkler, Faris and Roberts demonstrate that, despite the presence of Russian bots, their effects are marginal as compared to effects of mainstream media companies on political trends \citep{benkler2018network}. 

Skepticism is also warranted when blaming others for U.S. problems, as the U.S. government has a long history of stoking fear of other nations or people to justify its own aims. Many wars and territorial expansions throughout history have been sold to the nation through fear of a ``rogue regime'' or ``backwards people.'' The original project of colonization was sold through the ``Doctrine of Discovery'' and later ``Manifest Destiny,'' painting Indigenous people as dangerous savages who need to be civilized \citep{dunbar2014indigenous}. The Jim Crow era justified repression against new freedmen by depicting them as sexual predators and criminals \citep{pilgrim2002jim}. The second World War was sold on anti-Japanese propaganda \citep{dower2012war}, and the Cold War justified political coups and U.S. interventions, all sold as measures to combat Communist enemies \citep{chomsky1994manufacturing}. Even in our contemporary period, the wars in Iraq and Afghanistan were based on targeting a ``rogue regime'' on what turned out to be false pretenses \citep{rampton2003weapons}.

Domestic corporations have also played a significant role in disinformation campaigns through U.S. history. In the early 20th century, business lobby campaigns spreading anti-Bolshevist propaganda were later revealed to be aimed at weakening the power of labor unions by arguing they were run by foreign Communists \citep{carey1997taking}. The tobacco industry's manipulations are now famous examples of corporate disinformation, as are current analogues in climate change denial campaigns waged by the fossil fuel industry \citep{cook2019america}. 

Media organizations who purport to be objective are inevitably subject to the biases and ideologies of those who compose them. Media companies themselves have goals: attract large audiences, make money if they are for-profit, or advance a mission if they are non-profit. If that subjectivity is grounded in flawed political ideology, the information environment mirror those problematic views.

Returning to the ``objective'' media of the past, away from social media, does not seem to be the solution to combating disinformation. Moreover, focusing on rogue individuals and ignoring the structural issues with a media ecosystem, including biases and incentives to lie, will not address the roots of the problem. A more complicated and politically informed view of the institutional contributions to disinformation is sorely needed, and can take cues from historical disinformation campaigns to see which aspects should be studied presently.

\section{An Alternative, Sociopolitical Paradigm}

The analysis thus far leaves a crucial question: how to fill in these gaps and shortcomings? Taking a view more motivated by social motivations, individuals' subjectivity and positionality, political economics, and historical disinformation, leads us to a more \emph{sociopolitical} paradigm. At each layer, a more encompassing view can be described, which naturally articulates a different paradigm that lends new directions for disinformation research.

\subsection{The Content Layer}

\subsubsection{Rational, and empathetic}

A more complicated view of truth does not throw out judgement based on rationality, but it must include other crucial elements that contribute to the disinformation problem. Notably, something already identified as prevalent in disinformation campaigns is the use of oppressive language which may not be syntactically false, but is detrimental to a deliberative, democratic society (e.g., ``Muslims are among us''). When certain groups or individuals are the subject of dehumanizing, oppressive discourse, and subsequently deprived of empathy, environments are created where out-groups and enemies reduce the capacity for rational and reasonable discourse, as immoral speech is sanctioned \citep{stanley2015propaganda,cikara2011intergroup}. This type of environment triggers psychological progresses in belief formation that undermine rational and reasonable discourse, allowing more disinformation to spread and be believed.

When judging information quality, it stands to reason that beyond being judged for facticity, it can be judged for its empathetic or oppressive quality. Research projects can be developed to detect and categorize dimensions of unempathetic or oppressive speech using machine learning and natural language processing techniques, a small shift from the extant truth detecting systems. More research could also be conducted to determine what types of language and communication strategies foster empathy between groups, a crucial task to be undertaken if hateful divisions are to be addressed.

\subsubsection{Interpreted within its context and subjectivity}

There are also several dimensions along which information should be interpreted to best understand its effect in sociopolitical context. One is the positionality and institutional incentive structures of media producers and their journalists. The positionality of media producers includes their identity characteristics -- race, gender, sexuality, ethnicity, age -- and also their position in the social system -- educational attainment, wealth, and more. Institutional incentives also shape information production, as media outlets have certain goals which shape their editorial behavior and which stories are deemed publishable or not. Some incentives may include the pursuit of profit, a political agenda based on interests of a managing board, the political beliefs of top executive and editors, or any ties with other governments, corporations, or political actors. The combination of incentives and positionality shape how information is communicated and what frames it adopts, which is crucial to understanding its messaging and how it interacts with media consumers.

These frames are then part of larger ideologies and worldviews that make certain information more believable. Stories that argue for carbon taxes as a solution to the climate crisis, for example, would be more likely to be believed by those whose worldviews include more individualistic or market-based views of the world and how change happens. They may be less likely to be believed by those whose worldview sees government intervention as a key agent of change. Being able to identify different prominent ideologies and see their presence in media information would help add depth to what are otherwise simply media statements.

Another growing area of interest in analyzing disinformation content is the heavy use of trust cues. Media does not just encode facts, but also judgements about notable individuals and groups: politicians, executives, organizations, political groups \citep{moran2021trust}. Trust in media is a complicated topic, and what it includes or excludes is not agreed on among media researchers \citep{fawzi2021concepts,stromback2020news}, but it is a common refrain that trust in media is declining. Regardless, undoubtedly, reporting includes statements about the hypocrisy, authenticity, or trustworthiness of notable political entities. Being able to identify these cues and interpret statements by political entities taking into account a media consumer's trust in them is important for properly analyzing discourse and disinformation.

All of these informational cues and framings then interact with the subjective worldview of the media consumer. These subjective epistemologies \citep{marwick2018people} determine whether individuals believe disinforming content, and should be able to produce arguments as to why they believed it. Individuals should be more likely to believe information that matches their positionality, their held ideologies and worldviews, and comes from those whom they trust. Interpreting content within this context allows analysis of disinformation to break past rationality judgements alone, and move to include more complicated elements of what media messages encode and their interactions with consumers. 


\subsection{The Individual Layer}

\subsubsection{Subjective epistemologies}

An alternative paradigm would argue that to understand why disinformation flourishes, one must understand the subjective worldviews and epistemic systems of reasoning that believers hold. What may seem obviously false to one subjective worldview may seem entirely believable to another. While this does not mean that all subjective worldviews are equally valid, it does mean that understanding why disinformation is believed requires understanding the subjectivities that allowed it to be believed.

Different dimensions may be put forward to describe subjective worldviews, including what other political stances individuals take, the underlying political and social ideological that most resonate with them, or the figures and organizations that they most trust. Similarly to the description of the content layer above, each media consumer possesses complicated evaluations of issues and political actors. Being able to adequately describe these, and find which are most salient in different disinformation campaigns, could go far toward understanding why individuals adopt beliefs that seem obviously false to others.

\subsubsection{Psychological and emotional reasoning}

There is also an entirely complicated layer of psychological interaction that occurs when individuals choose whether or not to believe disinformation. Work on motivated cognition makes clear that there are several categories of psychological processes that affect what is believable or not: epistemic, existential, and relational motivation \citep{jost2003political,jost2009political}. For an example, climate denial may play to one's motivation to believe something that mitigates their anxiety about death. Social and relational motivations may also lead one to believe and disbelieve information according to whether it allows them to maintain status in a social group.

This consideration, in general, makes clear that what many may consider ``emotional'' reasoning is actually a large part of the belief process. In contrast to the classical rationalist view of the dominant paradigm, emotion and its role in certain disinformation narratives must be incorporated into analysis of belief.

The one area of emotional reasoning that has been widely adopted in the dominant paradigm is the consideration of cognitive dissonance reduction \citep{porot2020science} and its role in selective exposure \citep{metzger2020cognitive}. Any emotional turmoil caused by disconfirming information may lead to pressures to reduce cognitive dissonance. In particular, information that challenges a media consumer's sense of self, their identity, has a strong chance of doing so \citep{stanley2015propaganda}. To that end, a more complicated understanding of individuals' identities \citep{van2018partisan} and how information challenges their sense of self is necessary to understand how disinformation is adopted.

\subsubsection{Measuring subjective worldviews}

It stand to reason that, in the same vein as judging content for its oppressive and empathetic characteristics, individual worldviews can be evaluated along the same lines. These worldviews should be able to be measured, as similar work has measured motivated reasoning dimensions \citep{brady2019ideological,jost2018ideological} and conspiracy thinking \citep{uscinski2020people} in individuals. Constructing studies that gauge individuals' tendencies towards dominant oppressive ideologies -- racism, sexism, classism, xenophobia, etc. -- could be an easy intervention to begin understanding underlying context that interacts with information.

Building on psychological frameworks that describe ideological and social-motivational dimensions of reasoning, tools and frameworks could be developed that then gauge the interaction between contextualized media content and subjective individual worldviews. Instead of conducting wide analyses of disinformation narratives as if everyone is interacting with the content in the same way, incorporating worldview interactions could add important nuance to how disinformation spreads through certain networks, and why some narratives become more popular than others. These tools could also comment on the longer arc of political thought as media producers and consumers interact and societies tend towards certain ideologies and styles of thinking that change over time.

\subsection{The Group Layer}

\subsubsection{Group worldviews and ideologies}

As in the same way that embracing subjective epistemologies and worldviews is important for the content and individual layer, so too is it important for the group layer. Groups often form around shared worldviews, so understanding the ideological and sociopolitical views of, for example, QAnon supporters \citep{uscinski2022getting}, is necessary to understand why the group forms. Moreover, the frameworks of motivated reasoning foreground the importance of understanding relational dynamics, which may keep individuals in certain worldviews for fear of losing social status or group membership \citep{jost2009political}.

Being able to more descriptively quantify group worldviews along ideological, motivational, or sociopolitical dimensions can lend crucial insight also into the differences and similarities between groups. Being able to understand typical Democrat and Republican worldviews may lend insight into why Covid mask-wearing was more widely adopted among Democrats than Republicans \citep{pewMaskDifferences2020}.

Even beyond differences, however, a significant intervention that could be made based on this type of understanding are based on the similarities between groups' worldviews. As polarization forms around certain issues and affectively between different groups, being able to show similarities in worldviews to both groups could go far to foster empathy and understanding necessary for reasonable political discourse. If all that is focused on is differences, democratic deliberation becomes difficult. Being able to quantify and explicitly demonstrate group similarities would be a strong intervention against this tendency.

\subsubsection{In- and out-group dynamics}

Another important consideration for groups under an alternative paradigm is acknowledging and considering the effects that arise when in-groups and out-groups are formed from disinformation campaigns. Similarly to trust cues, certain groups of people are often disparaged by disinformation, leading to the withdrawal of empathy for them and political consequences for entire groups \citep{stanley2015propaganda,stanley2020fascism}. These dynamics can be very dangerous, as they have been identified as part of the core of what Stanley defines as fascist politics -- those which often precede genocide and ethnic cleansing.

In general, social psychology research has demonstrated that once in- and out-groups are established, typical morality judgements change \citep{cikara2011intergroup}. Immoral behavior towards out-groups becomes justifiable in a way that is not common when no group divisions are present. This type of group division also prohibits the reasonable interpretation of information that comes from an out-group \citep{stanley2015propaganda}, which may actually be key to forming the most rational opinion. If Democrats and Republicans become salient in- and out-groups, then each will tend to believe only what their in-group says, even if there are legitimate claims from the out-group. 

Disinformation research should strive to identify these groups as they form, and assess the belief of different campaigns within the context of group dynamics. Especially when considering disinformation that leads to immoral behavior and violence, as is common with campaigns that create support for wars, messaging that draws sharp ``us'' and ``them'' lines must be made salient when analyzing political messaging.

\subsection{The Institutional Layer}

\subsubsection{Institutional motivations and positionality}

The institutional layer also requires contextualizing by describing the motivations and positionality of different institutional actors. Media companies are staffed by certain individuals who have their own positionality. Moreover, media organizations follow certain mandates, whether those be to make profit for shareholders, to bring in advertisers, or to advance a mission.

Analysis of disinformation under this alternative paradigm requires that political economy be taken into consideration. For example, it must be acknowledged that a government will communicate information that helps advance its own interests, whether those be domestic or foreign policy, and should not be expected to be entirely truthful. An example is the Iraq War, justified by the U.S. government and media who followed it as a fight for democracy when it was later revealed to be a fight for control of oil and political territory \citep{rampton2003weapons}. Moreover, corporations also lie or bend the truth if it helps them further their goals. Fossil fuel corporations lied profusely in order to maintain their business prospects even though they knew they were contributing to global warming \citep{cook2019america}. Taking into account these goals will help identify disinformation as such.

\subsubsection{Corporate and U.S. government disinformation}

To these ends, disinformation study would benefit tremendously from the creation of new data sets and analyses that comment on corporate disinformation and U.S. government disinformation. The dominant paradigm concerns itself more with foreign actors and their attempts at influencing U.S. politics, but at the end of the day, large corporations and a domestic government have far more direct influence on the beliefs and attitudes of their consumers and citizens.

This represents an area of research that could spawn a plethora of new studies and digital artifacts, creating new corpora for further study. Moreover, comparative studies could be done to analyze the particular flavor of U.S. or corporate disinformation in comparison to the large body of work that has thus far been compiled on disinformation from foreign actors like Russia, China, or Iran.

\subsubsection{Historical analysis}

Combined with taking into account political economy, disinformation studies must include histories of both the actors perpetrating disinformation, and also general histories of disinformation. For the former, knowing, for example, that the U.S. government has a history of disinforming its population for the sake of getting involved in conflicts, is informative towards understanding present informational campaigns. Knowing that the rhetoric of U.S. corporations in the early 20th century was premised on Social Darwinism, justifying their monopolies by saying it was survival of the fittest, helps understand the rhetoric of disinforming corporations today by prompting analysts to look for other social logics that justify corporate behavior.

Additionally, taking into account general histories of disinformation can be generative, giving perspective to current circumstances that may seem inexplicable. Recounting historical disinformation campaigns waged in times of great social inequity \citep{carey1997taking}, or when trust in institutions was low \citep{putnam2000bowling}, can offer clues to why disinformation may seem more rampant today. Even certain historical examples of disinformation can be useful, as their tactics often mirror those of contemporary campaigns.

\subsection{A Holistic, Complex View}

In sum, the sociopolitical paradigm of disinformation addresses key parts of the problem that the dominant paradigm does not. It also has a coherence to the interplay of its various layers that the dominant paradigm does not. In each layer, it foregrounds the importance of analyzing information and those interacting with it along social and political lines: what are their positions in society, what are their motivations? This core distinction from the rationalist dominant paradigm, assuming that objective truth and rational deliberation is the guiding force of information adoption and spread, gives the sociopolitical paradigm a complexity and richness that should allow for more nuanced and accurate descriptions of disinformation events.

Beyond offering a more nuanced analysis of information itself, the sociopolitical paradigm also offers strengths in that its intertwining explanations at each level -- described in the language of position and motivation -- relate each layer back to the larger society. Thus, this paradigm does not view disinformation as solely a technological product, or one that is somehow divorced from the larger social conditions. Rather, it can only be explained by the social conditions \citep{koltai2022addressing}. To give an example, as inequality in society increases, this puts stress on individuals who are struggling more, which may make them psychologically prone to fear of loss \citep{jost2003political}, which can be played on by disinformation campaigns looking to create scapegoats to blame for poor social outcomes \citep{stanley2015propaganda} using racism, xenophobia, or classism. The subsequent racist or xenophobic messaging then creates racist or xenophobic worldviews in individuals that can be detected by analysis, and which can be further played to by new disinformation campaigns. This logic is more coherent than what is sometimes offered by the dominant paradigm, identifying the disinformation problem in several disparate, disjoint sources or phenomena.

Another strength of this alternative paradigm is that it can be used to build arguments that speak to a normative view of media and political communication, something that has been missing from the field \citep{anderson2021propaganda}. When evaluating ideologies of individuals, oppressive or empathetic cues in information, and institutional behaviors, the results of disinformation campaigns versus productive information campaigns can be explained and evaluated. In this sense, a vision of what a healthy information ecosystem looks like, or what beneficial political discourse contains, can be constructed from the results of research efforts.

None of this is to say that aspects of the dominant paradigm are not useful. Some pieces of the dominant view, particularly in its analysis of the new technological affordances created by the internet and social media, as well as their consequences, should be maintained as part of the problem. Likewise its pursuit of the most factual information available should not be abandoned, but simply contextualized as relative, not absolute truth. The sociopolitical paradigm simply adds key context, without which the technological and epistemological analysis falls short.

There is ample opportunity to conduct studies that work within the sociopolitical paradigm of disinformation, reporting on the content, individual, group, and institutional layers and their interactions with each other. To truly describe disinformation in its complexity, these should work side-by-side with studies that are investigating technical and logical aspects of the problem.

\section{Conclusion}

Though the effects of disinformation are widely felt, the mechanisms behind the problem are less clearly understood. In widely cited research and mainstream media discourse, U.S. disinformation is described as a failing of irrational or dumb individuals within echo-chambers, facilitated by the proliferation of social media and manipulations by foreign nations and rogue individuals. This dominant view argues for interventions and research in the form of fact-checking, a return to objective media, or nudging individuals towards truth through digital platforms.

This dominant view can be challenged by an alternative paradigm that describes the problem with more sociopolitical context. Though truth and falsehood are important, much disinformation relies on ideological, emotional, or trust cues to manipulate belief. Individuals' worldviews and rationalities are subjective, so what seems rational to them may seem irrational to an outside observer. Discourse between drastically different worldviews prohibits belief update, even if one view is more logical or rational. The worst disinformation plays on worldviews that include oppressive, empathy-reducing logics and ideologies, or play to in- and out-group narratives. This view also recognizes that many institutions have incentives that can lead them to disinform, whether they be profit or political power. A historical view of U.S. disinformation reveals frequent manipulation by corporations, political parties, and the government for these reasons.

With a more expansive and explanatory sociopolitical paradigm, there is a rich opportunity for a new wave of disinformation studies that speak to the contextual nature of information. Some examples are motivated -- including computational and psychological studies that structure news data within sociopolitical context, models that predict interactions between contextual information and individual or group worldviews, and studies of institutional actors who have previously been less examined -- but there is much room for development by the research community. Integrating this more holistic view of disinformation will allow for much more nuanced understanding of the current state of the world, as well as imagination of interventions that could have far reaching effects for moving society towards more productive discourse devoid of widespread disinformation. 

\bibliographystyle{abbrvnat}
\setcitestyle{authoryear,open={(},close={)}}

\end{document}